\documentclass[english,aps,preprintnumbers,amsmath,amssymb,nofootinbib,twocolumn,notitlepage,10pt,prd]{revtex4-1}

\usepackage[utf8]{inputenc}
\usepackage[T1]{fontenc}

\usepackage{ifthen}
\usepackage{forloop}
\usepackage[colorlinks]{hyperref}
\usepackage[english]{babel}
\usepackage{graphicx}

\usepackage{amsmath,amssymb,amsfonts}

\usepackage{mathrsfs}
\usepackage{bm}
\usepackage{bbm}

\usepackage{tensor}

\usepackage{slashed}

\usepackage{lipsum}
\usepackage{xcolor}

\let\oldFootnote\footnote
\newcommand\nextToken\relax

\renewcommand\footnote[1]{%
    \oldFootnote{#1}\futurelet\nextToken\isFootnote}

\newcommand\isFootnote{%
    \ifx\footnote\nextToken\textsuperscript{,}\fi}

\DeclareMathOperator*{\ext}{sup}

\newcommand{\euler}{\mathrm{e}}

\newcommand{\Tr}{\operatorname{Tr}}

\newcommand{\mcC}{\ensuremath{\mathcal{C}}}
\newcommand{\mcD}{\ensuremath{\mathcal{D}}}

\newcommand{\mcL}{\ensuremath{\mathcal{L}}}

\newcommand{\mcN}{\ensuremath{\mathcal{N}}}

\newcommand{\mrd}{\ensuremath{\mathrm{d}}}

\makeatother

\begin{document}

\title{
Renormalized Functional Renormalization Group
}
\author{Stefan Lippoldt}
\affiliation{Institut f\"ur Theoretische Physik, Universit\"at Heidelberg, 
Philosophenweg 16, 69120 Heidelberg, Germany}                                        

\begin{abstract}

We construct a new version of the effective average action together with its flow equation.
The construction entails in particular the consistency of fluctuation field and background field
equations of motion, even for finite renormalization group scales.
Here we focus on the quantum gravity application, while the generalization of this idea
to gauge theories is obvious.
Our approach has immediate impact on the background field approximation, which is
the most prominent approximation scheme within the asymptotic safety scenario.
We outline the calculation of quantum gravity observables from first principles using
the new effective average action.

\end{abstract}

\maketitle

\section{Introduction}
\label{sec:intro}

After more than 100 years of Einsteins theory of general relativity the search for the quantum theory of gravity
is still one of the most important open problems in theoretical physics.
Many different approaches are trying to shed light on this problem from various perspectives.
Ultimately the fate of the various quantum gravity models is decided by experiments.
Therefore, all these approaches have to make pre- or postdictions at some point.
In this work we outline a practical method to calculate observables within the asymptotic safety
scenario for quantum gravity \cite{Weinberg:1980gg, Reuter:1996cp}.

To calculate observables of any theory of quantum gravity the expectation value
$\langle \tilde{g} \rangle$ of the metric is of particular interest.%
\footnote{%
We use the tilde in $\langle \tilde{g} \rangle$ to indicate,
that the $\tilde{g}$ is not a fixed metric, but the integration variable
within a path integral,
$\langle \tilde{g} \rangle \sim \int \! \mcD \tilde{g} \, \tilde{g} \, \euler^{-S_{\rm cl}[\tilde{g}]}$,
where $S_{\rm cl}$ is the diffeomorphism invariant classical action of quantum gravity.
}
In order to derive $\langle \tilde{g} \rangle$ from first principles
one has to solve the quantum equations of motion,
\begin{align}
 0 = \left. \frac{\delta \Gamma[g]}{\delta g} \right|_{g = \langle \tilde{g} \rangle},
\end{align}
where $\Gamma$ is the quantum effective action of gravity.
Observables are then derived by combining the on-shell $n$-point correlators,
$\Gamma^{(n)}[\langle \tilde{g} \rangle]$, into gauge invariant objects.
Doing this for quantum gravity one can analyze the curvature invariants inside a black hole,
to see what happens to the singularity, or one can study the time evolution of the metric
to investigate cosmological inflation \cite{Falls:2010he, Bonanno:2017pkg}.

One of the key goals of asymptotic safety is the derivation of the effective average action $\Gamma_{k}$,
which in the physical limit, $k \to 0$, approaches the quantum effective action,
$\Gamma_{k} \stackrel{k \to 0}{\longrightarrow} \Gamma$.
This derivation involves two main steps: one first needs to find the eponymous asymptotically safe fixed point
in the ultraviolet and in the second step one has to integrate the renormalization group flow down to the infrared
leading to $\Gamma$.
Under certain circumstances, e.g., for single scale problems,
one can use $\Gamma_{k}$ also for finite scales $k$ instead of integrating down completely.
The reason is, that in these cases the flow of the relevant correlators essentially stops shortly below the
present physical scale.

In the literature the by far most used approximation scheme is the background field approximation,
\cite{Reuter:1996cp, Gies:2006wv}.
The main advantages are its seemingly manifest diffeomorphism invariance and a manageable amount of necessary calculations.
So it was, e.g., possible to demonstrate that asymptotic safety actually is safe against the famous Goroff-Sagnotti counter
term, which was believed to spoil the ultraviolet fixed point, as it marks the failure of perturbative quantum gravity,
\cite{Gies:2016con}.
Despite the widespread use of the background field approximation, it still is an approximation.
By now there are several works pointing towards quite some tension between
proper fluctuation field calculations and the background field approximation,
cf.\ \cite{Bridle:2013sra}, \cite{Benedetti:2009rx, Christiansen:2016sjn},
\cite{Dona:2013qba, Meibohm:2015twa, Christiansen:2017cxa} and \cite{Eichhorn:2018akn}.
These discrepancies are expected at least for nonzero renormalization group scales $k$,
due to nontrivial split Ward identities, cf.\ equation \eqref{eq:WTI} and \cite{Freire:2000bq}.
Unfortunately, as discussed above, this is exactly what one would like to do:
use $\Gamma_{k}$ for finite scales $k$.

The reason why one actually has to track two separate fields is as follows.
One can artificially split the full metric $g$ into a background metric $\bar{g}$
and a fluctuation field $h$ by $g = \bar{g} + h$.%
\footnote{%
We focus here on the linear split, while other splits can be discussed similarly.
}
This simple split gets broken,
due to the need for gauge fixing and regularization.
Therefore naively $\Gamma_{k}$ is a function of $\bar{g}$ and $h$ separately,
even though these fields are actually related by the nontrivial split Ward identities \eqref{eq:WTI}.
This is nothing special to gravity, the same idea applies to gauge theories in the background field formulation,
cf.\ \cite{Freire:2000bq, Pawlowski:2005xe, Gies:2006wv}.
However there, other than in gravity, one does not have to introduce a background.

In the literature there are essentially two kinds of approaches
trying to investigate the separate dependence of $\Gamma_{k}$ on $\bar{g}$ and $h$.
The one kind deals with solving the split Ward identities in order to formulate the theory in terms of only one field,
\cite{Dietz:2015owa, Safari:2015dva, Morris:2016nda, Labus:2016lkh, Safari:2016gtj, Morris:2016spn,
Percacci:2016arh, Ohta:2017dsq, Nieto:2017ddk}.
The other uses the fact, that if these split Ward identities are satisfied at a single renormalization group scale,
then they are satisfied for all scales, if the flow is carried out in an exact manner.
Therefore, one can in a first step forget about the identities and simply study the theory involving both fields,
while at the end only making sure that the split Ward identities are satisfied in the infrared,
\cite{Manrique:2009uh, Manrique:2010am, Christiansen:2014raa, Becker:2014qya,
Christiansen:2015rva, Denz:2016qks, Knorr:2017fus, Christiansen:2017bsy}.
Let us mention that there are also some geometric approaches, trying to deal with the separate dependence
on the background and the fluctuation field in an explicitly gauge invariant manner
\cite{Branchina:2003ek, Pawlowski:2003sk, Donkin:2012ud, Wetterich:2016ewc, Wetterich:2017aoy}.

In this letter we follow a new direction en route to the above problem.
We present a simple modification of the standard effective average action in section \ref{sec:Renormalized_FRG}.
It guarantees that the equations of motion for the background
field are compatible with the true quantum equations of motion of the full quantum field.
Furthermore, this modification leads to several nice properties of the effective average action,
cf.\ section \ref{sec:Properties_of_Gamma_hat}.
In particular we find a simple relation between the pure background effective average action
and the remainder, which then also involves the fluctuation field.
That is to say, a certain subset of the split Ward identities can be cast in a comparatively simple form.
In this way we help to improve the understanding in both directions,
solving the split Ward identities on the one hand
and the study of both fields independently on the other hand.

The remainder of this section \ref{sec:intro} can be safely skipped by readers
not familiar with the functional renormalization group.
We consider here the calculation of the expectation value of the metric, $\langle \tilde{g} \rangle_{k}$,
to illustrate how the modification of the effective average action works.
As $\langle \tilde{g} \rangle_{k}$ itself is not an observable,
explicit calculations will depend on the choice of the background.
However, it is to be expected that one gets a good estimate if one chooses the background such,
that the expectation value of the fluctuation field vanishes \cite{Reuter:1997gx}.
This then corresponds to an expansion about the solution of the equations of motion.
In this case the background field and the expectation value of the metric are identical.
Hence, the defining equation for $\langle \tilde{g} \rangle_{k}$ is
\begin{align}
 \label{eq:vev_metric_Gamma_Fluc}
 0 = \left. \frac{\delta \Gamma_{k}[h; \bar{g}]}{\delta h}
 \right|_{\substack{h = 0 \hspace{0.4cm} \\ \bar{g} = \langle \tilde{g} \rangle_{k}}}.
\end{align}
The new idea is to modify the effective average action $\Gamma_{k}$,
such that the equations of motion for $h$ and $\bar{g}$ are compatible at finite renormalization
group scales $k$, without changing the quantum physics, i.e., the dynamics of $h$.
We show in section \ref{sec:Renormalized_FRG}, that this can be achieved by defining $\hat{\Gamma}_{k}$
as the Legendre transform,%
\footnote{%
Here and in the following we mostly suppress the Faddeev-Popov ghosts as they are not important for our discussion.
}
\begin{align}
 \label{eq:Def_Gamma_hat_Legendre_Trafo}
 \hat{\Gamma}_{k}[h ; \bar{g}]
 = \ext\limits_{J} \big( J \cdot h - \hat{W}_{k}[J;\bar{g}] \big) - \Delta S_{k}[h;\bar{g}],
\end{align}
where $\hat{W}_{k}$ is a properly normalized Schwinger functional,
\begin{align}
 \label{eq:Def_Schwinger_hat}
 \hat{W}_{k}[J;\bar{g}] = \ln \frac{Z_{k}[J;\bar{g}]}{Z_{k}[0;\bar{g}]},
\end{align}
with the partition function $Z_{k}$, cf.\ equation \eqref{eq:Def_partition_function}.
The difference between $\Gamma_{k}$ and $\hat{\Gamma}_{k}$ is the normalization $Z_{k}[0;\bar{g}]$
in the definition of the Schwinger functional \eqref{eq:Def_Schwinger_hat}.
As $\hat{\Gamma}_{k}$ is built up entirely of elements already present in the standard formulation,
the new flow equation is rather similar to the standard one, cf.\ equation \eqref{eq:new_flow_equ}.

It is important to note, that the normalization $Z_{k}[0;\bar{g}]$ in equation \eqref{eq:Def_Schwinger_hat}
only depends on the background.
Therefore it does not have an impact on the fluctuation correlators containing the physics,
\begin{align}
 \label{eq:h_deriv_Gamma_Gamma_hat}
 \frac{\delta \Gamma_{k}[h; \bar{g}]}{\delta h}
 = \frac{\delta \hat{\Gamma}_{k}[h; \bar{g}]}{\delta h}.
\end{align}
One can check that this additional background term ensures that the solution,
$\langle \tilde{g} \rangle_{k}$, of equation \eqref{eq:vev_metric_Gamma_Fluc} also is a solution
of the analogous equation for the background field,
\begin{align}
 \label{eq:vev_metric_Gamma_Back}
 0 = \left. \frac{\delta \hat{\Gamma}_{k}[h; \bar{g}]}{\delta \bar{g}}
 \right|_{\substack{h = 0 \hspace{0.4cm} \\ \bar{g} = \langle \tilde{g} \rangle_{k}}}.
\end{align}
Therefore, instead of calculating the expectation value of the metric using the fluctuation field,
equation \eqref{eq:vev_metric_Gamma_Fluc},
we can equivalently use the background field, equation \eqref{eq:vev_metric_Gamma_Back}.
Thus in future work one can improve the background field approximation by using $\hat{\Gamma}_{k}$.

\section{Renormalized FRG}
\label{sec:Renormalized_FRG}

The asymptotic safety scenario relies on the idea of an interacting ultraviolet fixed point for quantum gravity.
Therefore a perturbative treatment is no option.
One way to investigate the properties of such an interacting fixed point is the study of the
effective average action, $\Gamma_{k}$, together with the exact flow equation, cf.\ \cite{Wetterich:1992yh, Reuter:1996cp},
\begin{align}
 \label{eq:flow_equ}
 \dot{\Gamma}_{k}[h;\bar{g}]
 = \frac{1}{2} \Tr \left[
 \big( \Gamma_{k}^{(2;0)}[h;\bar{g}]
 + R_{k}[\bar{g}] \big)^{-1} \dot{R}_{k}[\bar{g}]
 \right].
\end{align}
This equation is an explicit implementation of Wilson's renormalization group idea of integrating out
the momenta shell-by-shell.
Due to the specific properties of the regulator $R_{k}$ the flow of $\Gamma_{k}$ is driven by modes
close to the renormalization group scale $k$.
For $k \to 0$ the regulator vanishes and the effective average action approaches the full quantum effective action,
$\Gamma_{k} \stackrel{k \to 0}{\longrightarrow} \Gamma$.

By definition the information about quantum physics is contained in the correlators of the fluctuation field,
while the background field is just a technical aid.
As discussed earlier the simple linear split, $g = \bar{g} + h$, is broken due to the presence of the gauge fixing,
$S_{\rm gf}$, the ghosts, $S_{\rm gh}$, and the regulator,
$\Delta S_{k}[h;\bar{g}] = \frac{1}{2} h \cdot R_{k}[\bar{g}] \cdot h$,
leading to the (modified) split Ward identities, cf.\ \cite{Freire:2000bq},
\begin{align}
 \label{eq:WTI}
 (\tfrac{\delta}{\delta h} {}& - \tfrac{\delta}{\delta \bar{g}} ) (\Gamma_{k} + \Delta S_{k})
 = \big\langle (\tfrac{\delta}{\delta \tilde{h}} - \tfrac{\delta}{\delta \bar{g}} )
 (S_{\rm gf} + S_{\rm gh} + \Delta S_{k}) \big\rangle^{J}_{\bar{g}}.
\end{align}
Here the $\langle \cdot \rangle^{J}_{\bar{g}}$ denotes the expectation value in presence of the sources, $J$,
and the background field, $\bar{g}$.
Due to the above identities the deviation of fluctuation and background field derivatives
is only determined by unphysical terms, i.e., the regulator and the gauge fixing.
Hence, the difference drops out when observables are calculated in the limit $k \to 0$.
Therefore, once we have $\Gamma[h;\bar{g}]$, we can restrict ourselves to the gauge invariant functional
$\Gamma[0;g]$ and use this to calculate all the observables.
This insight lies at the heart of the background field approximation.
Furthermore, one can show that the right hand side of equation \eqref{eq:WTI} essentially corresponds to the background
field derivative of the partition function $Z_{k}$,
\begin{align}
 \big\langle (\tfrac{\delta}{\delta \tilde{h}} - \tfrac{\delta}{\delta \bar{g}} )
 (S_{\rm gf} + S_{\rm gh} + \Delta S_{k}) \big\rangle^{J}_{\bar{g}}
 = J + \frac{\delta \ln Z_{k}[J;\bar{g}]}{\delta \bar{g}}.
\end{align}
The partition function is defined as
\begin{align}
 \label{eq:Def_partition_function}
 {}& Z_{k}[J;\bar{g}] = \! \int \!\! \mcD \tilde{h}
 \, \Delta_{\rm FP}[\tilde{h};\bar{g}] \,
 \euler^{- S_{\rm cl}[\bar{g} + \tilde{h}] - S_{\rm gf}[\tilde{h};\bar{g}]
 - \Delta S_{k}[\tilde{h};\bar{g}] + J \cdot \tilde{h}},
\end{align}
where $S_{\rm cl}$ is the diffeomorphism invariant classical action,
and $\Delta_{\rm FP}$ is the Faddeev-Popov determinant.
This partition function is background independent when evaluated at $k=0$ for vanishing sources, $J=0$.%
\footnote{%
As we discuss in section \ref{sec:Properties_of_Gamma_hat} this is not quite true for quantum gravity.
The gauge-fixing procedure introduces a background dependent normalization into the path integral,
leading to a slightly different definition of a background independent partition function,
cf.\ equation \eqref{eq:Def_background_independent_partition_function}.
}
By definition the sources, $J$, vanish when they are evaluated on-shell.
Therefore we find the anticipated compatibility of the equations of motion for the background and the fluctuation field
in the limit $k \to 0$, \cite{Reinosa:2014ooa, Christiansen:2017bsy},
\begin{align}
 0 = \left. \frac{\delta \Gamma[h;\bar{g}]}{\delta h}
 \right|_{\substack{h = 0 \hspace{0.25cm} \\ \bar{g} = \langle \tilde{g} \rangle}}
 = \left. \frac{\delta \Gamma[h;\bar{g}]}{\delta \bar{g}}
 \right|_{\substack{h = 0 \hspace{0.25cm} \\ \bar{g} = \langle \tilde{g} \rangle}}.
\end{align}
However, from equation \eqref{eq:WTI} it is apparent, that the compatibility cannot hold for finite values of $k$.
The problem is the unavoidable separate fluctuation field and background field dependence of the regulator $\Delta S_{k}$.
We propose to bypass this problem, by modifying the background field dependence of the effective average action,
while leaving the fluctuation field dependence untouched.
The usual effective average action is defined as the Legendre transform of the Schwinger functional,
\begin{align}
 \Gamma_{k}[h ; \bar{g}]
 = \ext\limits_{J} \big( J \cdot h - \ln Z_{k}[J;\bar{g}] \big) - \Delta S_{k}[h;\bar{g}].
\end{align}
The crucial step now is to use the normalized Schwinger functional $\hat{W}_{k}$, cf.\ equation \eqref{eq:Def_Schwinger_hat}.
We immediately see, that we only added a background dependent term.
Hence, if we calculate an observable using the fluctuation field, we take derivatives of $\hat{W}_{k}$
with respect to the source $J$ and this additional piece immediately drops out.
However, if we use the background field to calculate an observable the second piece contributes,
while still dropping out for $k \to 0$, due to the background invariance of the partition function.
Using this normalized Schwinger functional leads to a normalized effective average action $\hat{\Gamma}_{k}$,
\begin{align}
 \label{eq:Def_Gamma_hat}
 \hat{\Gamma}_{k}[h;\bar{g}]
 = \Gamma_{k}[h;\bar{g}] - \Gamma_{k}\big[ h_{k}[\bar{g}] ; \bar{g} \big]
 - \Delta S_{k}\big[ h_{k}[\bar{g}] ; \bar{g} \big],
\end{align}
where the background dependent field $h_{k}[\bar{g}]$ is the solution to
the quantum equations of motion at some finite renormalization group scale $k$,
\begin{align}
 \label{eq:quantum_EOM}
 0 = \left. \frac{\delta}{\delta h} \big( \hat{\Gamma}_{k}[h;\bar{g}]
 + \Delta S_{k}[h;\bar{g}] \big) \right|_{h = h_{k}[\bar{g}]}.
\end{align}
The presence of the regulator, $\Delta S_{k}$, ensures that the sources appearing
in the partition function also vanish for finite values of $k$ and
for arbitrary backgrounds $\bar{g}$.%
\footnote{%
In the physical limit, $k \to 0$, this term vanishes anyway.
}

The above definitions lead to the split Ward identities in terms of $\hat{\Gamma}_{k}$,
\begin{align}
 \notag
 (\tfrac{\delta}{\delta h} \! - \! \tfrac{\delta}{\delta \bar{g}} ) (\hat{\Gamma}_{k} \! + \! \Delta S_{k})
 ={}& \big\langle (\tfrac{\delta}{\delta \tilde{h}} \! - \! \tfrac{\delta}{\delta \bar{g}} )
 (S_{\rm gf} \! + \! S_{\rm gh} \! + \! \Delta S_{k}) \big\rangle^{J}_{\bar{g}}
 \\
 {}& - \big\langle (\tfrac{\delta}{\delta \tilde{h}} \! - \! \tfrac{\delta}{\delta \bar{g}} )
 (S_{\rm gf} \! + \! S_{\rm gh} \! + \! \Delta S_{k}) \big\rangle^{0}_{\bar{g}}.
\end{align}
It is now straightforward to see that the equations of motion of the fluctuation field
and the background field are compatible also for finite $k$, since $J$ vanishes on-shell,
\begin{align}
 0 ={}& \left. \frac{\delta}{\delta \bar{g}} \big( \hat{\Gamma}_{k}[h;\bar{g}]
 + \Delta S_{k}[h;\bar{g}] \big) \right|_{h = h_{k}[\bar{g}]}.
\end{align}
Considering the quantum equations of motion \eqref{eq:quantum_EOM} for finite $k$,
one might wonder about the impact of the regulator.
As the regulator is unphysical, it seems desirable to eliminate it from this equation.
Since it is quadratic in the fluctuation field, this thought suggests a preferred background,
namely $\langle \tilde{g} \rangle_{k}$, cf.\ equation \eqref{eq:vev_metric_Gamma_Fluc}.
This is the background for which the solution to the quantum equations of motion,
$h_{k}[\bar{g}]$, vanishes, leading us to equation \eqref{eq:vev_metric_Gamma_Back}.

Finally let us look at the flow equation for $\hat{\Gamma}_{k}$.
We already established equation \eqref{eq:Def_Gamma_hat}.
Hence, we get the flow of $\hat{\Gamma}_{k}$ by taking the logarithmic $k$
derivative and then use the standard flow equation \eqref{eq:flow_equ}
together with the on-shell condition \eqref{eq:quantum_EOM}
while also noting \eqref{eq:h_deriv_Gamma_Gamma_hat},
\begin{align}
 \notag
 k \partial_{k} \hat{\Gamma}_{k}[h;\bar{g}]
 ={}& \frac{1}{2} \Tr \left[
 \big( \hat{\Gamma}_{k}^{(2;0)}[h;\bar{g}]
 + R_{k}[\bar{g}] \big)^{-1} \dot{R}_{k}[\bar{g}]
 \right]
 \\ \notag
 {}& - \frac{1}{2} \Tr \left[
 \big( \hat{\Gamma}_{k}^{(2;0)}[h_{k}[\bar{g}];\bar{g}]
 + R_{k}[\bar{g}] \big)^{-1} \dot{R}_{k}[\bar{g}]
 \right]
 \\ \label{eq:new_flow_equ}
 {}& - \frac{1}{2} h_{k}[\bar{g}] \cdot \dot{R}_{k}[\bar{g}] \cdot h_{k}[\bar{g}].
\end{align}
Here the first line is the standard flow equation and is the only piece that depends on the fluctuation field.
The second and third lines are the new terms, accounting for the compatibility of fluctuation field and background field
equations of motion.

\section{Features of \texorpdfstring{$\hat{\Gamma}_{k}$}{Gamma hat}}
\label{sec:Properties_of_Gamma_hat}

The newly achieved compatibility leads to a plethora of very nice properties
of the effective average action, \cite{Lippoldt:2018a}.
As a first example we consider the limit $k \to \infty$.
In some cases, due to the divergence of the regulator, $R_{k} \stackrel{k \to \infty}{\longrightarrow} \infty$,
the method of steepest descent is applicable to evaluate the path integral
of the partition function.
We note that the normalization $Z_{k}[0;\bar{g}]$ in equation \eqref{eq:Def_Schwinger_hat}
leads to a well defined $k \to \infty$ limit of $\hat{\Gamma}_{k}$. 
The problem for $\Gamma_{k}$ is that there appears an ill defined and background dependent prefactor
in the partition function, $Z_{k \to \infty} \sim \frac{1}{\sqrt{\det R_{k}}}$.
This cumbersome prefactor exactly cancels out for $\hat{\Gamma}_{k}$.

The normalization $Z_{k}[0;\bar{g}]$ has another appreciated benefit.
It implies that every possible purely background dependent modification of the path integral measure
automatically drops out in equation \eqref{eq:Def_Schwinger_hat}.
This is particularly useful for quantum gravity, as the standard inclusion of the gauge fixing
\eqref{eq:Def_partition_function} only preserves the background invariance of the partition function up to
an ultralocal normalization.
To see this, we have a look at the Faddeev-Popov procedure.
After choosing a gauge fixing condition $F[h;\bar{g}]$ we introduce the Faddeev-Popov determinant,
$\Delta_{\rm FP}[h;\bar{g}]$, by integrating the functional delta distribution, $\delta(F[h^{\varepsilon};\bar{g}] - \mcC)$,
over the gauge group,
$1 = \Delta_{\rm FP}[h;\bar{g}] \int \! \mcD \varepsilon \, \delta(F[h^{\varepsilon};\bar{g}] - \mcC)$.
Here $\mcC$ is an arbitrary vector field and $h^{\varepsilon}$ is the gauge transformed fluctuation field,
where the background field is held constant.
For an infinitesimal diffeomorphism it reads $h^{\varepsilon} = h + \mcL_{\varepsilon}(\bar{g} + h)$,
where $\mcL_{\varepsilon}$ is the Lie derivative in $\varepsilon$ direction.
This factor $1$ is then inserted into the path integral and one further integrates over the arbitrary vector field
$\mcC$ with an exponential weight,
$\frac{1}{\mcN[\bar{g}]}
\euler^{- \frac{1}{\alpha} \! \int \!\! \mrd^{d} x \sqrt{ \bar{g}(x) } \, \mcC^{\mu}(x) \bar{g}_{\mu \nu}(x) \mcC^{\nu}(x)}$,
in order to transform the delta distribution into the standard gauge-fixing action.
The integration introduces the background field dependent ultralocal normalization $\mcN[\bar{g}]$.
This normalization is necessary as the exponential weight is designed to ensure the background-gauge invariance,
and thus needs to depend on the background metric, leading to
\begin{align}
 \mcN[\bar{g}]
 = \! \int \!\! \mcD \mcC \
 \euler^{- \frac{1}{\alpha} \! \int \!\! \mrd^{d} x \sqrt{ \bar{g}(x) } \, \mcC^{\mu}(x) \bar{g}_{\mu \nu}(x) \mcC^{\nu}(x)}.
\end{align}
Therefore, the properly gauge fixed and background invariant partition function reads,
\begin{align}
 \label{eq:Def_background_independent_partition_function}
 Z
 = \frac{1}{\mcN[\bar{g}]} \int \!\! \mcD \tilde{h} \, \Delta_{\rm FP}[\tilde{h}; \bar{g}] \,
 \euler^{- S_{\rm cl}[\bar{g} + \tilde{h}] - S_{\rm gf}[\tilde{h};\bar{g}]},
\end{align}
where the gauge-fixing action is given by
\begin{align}
 S_{\rm gf}[h;\bar{g}]
 = \frac{1}{\alpha} \! \int \!\! \mrd^{d} x \sqrt{ \bar{g}(x) } \,
 F[h;\bar{g}]^{\mu}(x) \, \bar{g}_{\mu \nu}(x) \, F[h;\bar{g}]^{\nu}(x).
\end{align}
We stress, that the normalization $\mcN[\bar{g}]$ is important to ensure the background invariance of the partition function,
but usually is neglected, cf.\ equation \eqref{eq:Def_partition_function}.
Such ultralocal terms are commonly disregarded, as they do not contribute, when one is using dimensional regularization.
However, how far this also applies to the functional renormalization group is not clear.
In particular it seems that the equations of motion for the fluctuation field and the background field in terms of $\Gamma$
can only be compatible, if the normalization $\mcN[\bar{g}]$ is taken into account.
Nevertheless, as discussed above, this is automatically taken care of
in the normalized effective average action $\hat{\Gamma}_{k}$.

Another nice property of the new effective average action is the proper on-shell normalization in the following sense.
One can show that the on-shell $\hat{\Gamma}_{k}$ has a simple form for all values of $k$,
\begin{align}
 \hat{\Gamma}_{k}\big[ h_{k}[\bar{g}] ; \bar{g} \big] + \Delta S_{k}\big[ h_{k}[\bar{g}] ; \bar{g} \big] = 0.
\end{align}
By taking repeated functional derivatives with respect to the background field,
one casts a certain subset of the split Ward identities in a particularly simple fom.
This subset relates the pure background on-shell $n$-point functions,
$\hat{\Gamma}_{k}^{(0;n)}\big[h_{k}[\bar{g}];\bar{g}]$,
to the mixed fluctuation and background $(r;s)$-point functions,
$\hat{\Gamma}_{k}^{(r;s)}\big[ h_{k}[\bar{g}] ; \bar{g} \big]$, with $r \in \{ 1, \ldots , n\}$
and $s \in \{ 0 , \ldots , n-r \}$.
Using this, one can establish an improved vertex expansion, around the on-shell configurations of the theory.
Hence, we can derive the flow of the vertices entering the calculation of observables,
namely the on-shell vertices, in contrast to the vertices at some arbitrary field configuration,
as in the standard vertex expansion.

Let us sketch, why the direct calculation of on-shell vertices is a very important task already for only
capturing the correct qualitative quantum behavior of gravity.
One of the central objects for deriving quantum effects is the quantum propagator
as it enters in all diagrams.
In particular points in theory space where it is enhanced usually correspond to strong quantum effects.
To be precise let us consider the transverse-traceless part of the graviton propagator for the Einstein-Hilbert action,
including a cosmological constant, evaluated on a maximally symmetric background,
$\frac{1}{\Delta - 2 \Lambda + \frac{2}{3} \bar{R}}$.
Here $\Delta$ is the covariant Laplacian, $\bar{R}$ is the Ricci scalar and $\Lambda$ is the cosmological constant.
When regularized in the functional renormalization group, we essentially replace $\Delta$ by $k^{2}$.
This leads to the well known pole of $\Lambda = \frac{k^{2}}{2}$ in theory space,
when we evaluate this propagator on a flat background.
In particular already coming close to this pole leads to a strong enhancement of quantum gravity effects.
However, if we now do the same analysis around the on-shell background, i.e., $\bar{R} = 4 \Lambda$, we find
$\frac{1}{\Delta + \frac{2}{3} \Lambda}$ for the propagator.
Thus, the position of the pole now is at $\Lambda = - \frac{3}{2} k^{2}$.
As the position has changed drastically, particularly even the sign has changed,
the qualitative behavior of the propagator is very sensitive in this respect.
This observation is of particular interest for a cosmologically viable renormalization group trajectory.
Due to the naive pole at $\Lambda = \frac{k^{2}}{2}$ it seems that we cannot connect the
ultraviolet fixed point to a finite positive cosmological constant within the background field approximation,
cf.\ \cite{Reuter:2001ag}.
However, we have argued that the location of this pole might be an artifact of not expanding
around the on-shell background, resolving this problem.

Another important point is again related to the propagator and in particular to gauge symmetry.
From standard quantum field theory textbooks we know, that the not gauge fixed propagator has zero modes.
In fact this is the practical reason we need to gauge fix.
Let us consider the inverse propagator of Einstein-Hilbert in flat space,
\begin{align}
 \notag
 S^{(2)}_{\rm EH}
 ={}&
 \frac{1}{32 \pi G_{\rm N}}
 \big(
 (p^{2} - 2 \Lambda) \, \Pi_{\rm TL}
 - (3 p^{2} - 2 \Lambda) \, \Pi_{\rm Tr}
 \\
 {}& \hspace{1.3cm}
 + \delta p p + p p \delta - 2 \, p \delta p
 \big),
\end{align}
where we used $\tensor{(\Pi_{\rm TL})}{_{\mu \nu}^{\rho \sigma}}
= \frac{1}{2} \delta_{\mu}^{\rho} \delta_{\nu}^{\sigma} + \frac{1}{2} \delta_{\mu}^{\sigma} \delta_{\nu}^{\rho}
- \frac{1}{4} \delta_{\mu \nu} \delta^{\rho \sigma}$,
$\tensor{(\Pi_{\rm Tr})}{_{\mu \nu}^{\rho \sigma}} = \frac{1}{4} \delta_{\mu \nu} \delta^{\rho \sigma}$,
$\tensor{(\delta p p)}{_{\mu \nu}^{\rho \sigma}} = \delta_{\mu \nu} p^{\rho} p^{\sigma}$,
$\tensor{(p p \delta)}{_{\mu \nu}^{\rho \sigma}} = p_{\mu} p_{\nu} \delta^{\rho \sigma}$,
and $\tensor{(p \delta p)}{_{\mu \nu}^{\rho \sigma}}
= \frac{1}{4} p_{\mu} \delta_{\nu}^{\rho} p^{\sigma} \! + \! \frac{1}{4} p_{\nu} \delta_{\mu}^{\rho} p^{\sigma}
\! + \! \frac{1}{4} p_{\mu} \delta_{\nu}^{\sigma} p^{\rho} \! + \! \frac{1}{4} p_{\nu} \delta_{\mu}^{\sigma} p^{\rho}$.
In order to derive the propagator we have to invert the above expression.
Naively we would expect that we cannot do this, due to the presence of zero modes related to the gauge symmetry.
However, an explicit computation reveals that there is no problem in deriving the inverse,
as long as $\Lambda \neq 0$,
\begin{align}
 \label{eq:EH_Propagator_Flat}
 \frac{1}{S^{(2)}_{\rm EH}}
 = \frac{32 \pi G_{\rm N}}{p^{2} - 2 \Lambda} \big( \Pi_{\rm TL} - \Pi_{\rm Tr} - \tfrac{1}{\Lambda} p \delta p \big).
\end{align}
To see why this is the case, we consider the gauge invariance of the Einstein-Hilbert action,
\begin{align}
 0 = S^{(1)}_{\rm EH}[g] \cdot \mcL_{\varepsilon} g.
\end{align}
By taking another functional derivative with respect to the metric we find
\begin{align}
 \label{eq:zero_modes_S(2)}
 0 = S^{(2)}_{\rm EH}[g] \cdot \mcL_{\varepsilon} g
+ S^{(1)}_{\rm EH}[g] \cdot \frac{\delta \mcL_{\varepsilon} g}{\delta g}.
\end{align}
As expected the zero mode of $S^{(2)}_{\rm EH}$ is in the gauge direction $\mcL_{\varepsilon} g$.
However, it only is a zero mode, if the second term in equation \eqref{eq:zero_modes_S(2)} vanishes.
This is the case, when we go on-shell.
As the flat background is no solution to the equations of motion for nonvanishing $\Lambda$,
we don't have a zero mode in $S^{(2)}_{\rm EH}$.
However, when switching off the cosmological constant, then the flat background is a solution, and we see a divergence
in the propagator \eqref{eq:EH_Propagator_Flat}.
Therefore, only the expansion around an on-shell background leads to the from gauge symmetry expected zero modes.
In conclusion: the normalized effective average action, $\hat{\Gamma}_{k}$, is perfectly suited for
quantum gravity calculations, as it introduces a certain on-shellness into the flow equation
by the presence of $h_{k}[\bar{g}]$ in equation \eqref{eq:new_flow_equ}.

\section{Outlook}
\label{sec:Outlook}

In this paper we have developed a normalized effective average action together with its flow equation.
It is designed such that the background field and the fluctuation field equations of motion are compatible
even for finite renormalization group scales.
Doing so we paved the way for improved calculations tackling various problems in quantum gravity.

A first application could be to study the impact of the new terms in the flow equation
within the background field approximation.
Using this one might be able to lift some of the observed tensions between the background field approximation
and the fluctuation field calculations.
At the same time, these improved calculations can be used to strengthen results already
derived in the background field approximation.
There a particular focus could lie in the expectation value of the metric, $\langle \tilde{g} \rangle_{k}$,
in order to study the quantum gravity effects on the singularity of a black hole or on inflation.

Furthermore, one can go beyond the background field approximation, and use this improved compatibility of
fluctuation field and background field dependence of $\hat{\Gamma}_{k}$.
In this direction, one can either try to use the simplified split Ward identities,
discussed in section \ref{sec:Properties_of_Gamma_hat}, in order to restrict the theory space,
or one can setup an assisted level-one improvement in a mixed fluctuation and background field calculation,
cf.\ \cite{Eichhorn:2018akn}.
Similarly one can make use of the on-shell background, $\langle \tilde{g} \rangle_{k}$,
in order to set up an on-shell vertex expansion.
The latter is of particular interest for quantum gravity as the nontrivial background
definitely plays a crucial role, already for the qualitative quantum behavior.

Finally let us note, that this approach of properly normalizing the effective average action
also works more generally, i.e., for every theory where a nontrivial background is of help.
Thus the benefit for gauge theories or theories with nontrivial vacua is immediate.
Particularly the vertex expansion about an on-shell background allows for the direct derivation
of the vertices entering in the calculation of observables and therefore constitutes
an important step for the functional renormalization group framework.

\acknowledgments

I thank J.~Pawlowski and M.~Reichert for discussions.

\bibliography{general_bib}

\end{document}